# SURVEYING ATTITUDINAL ALIGNMENT BETWEEN LARGE LANGUAGE MODELS VS. HUMANS TOWARDS 17 SUSTAINABLE DEVELOPMENT GOALS

**Qingyang Wu**[*]
University of California, Los Angeles
qw@ucla.edu

**Ying Xu**[*]
Washington University in St. Louis
x.ying1@wustl.edu

**Tingsong Xiao**
University of Florida
xiaotingsong@ufl.edu

**Yunze Xiao**
Carnegie Mellon University
yunzex@andrew.cmu.edu

**Yitong Li**
Stanford University
yitongli@stanford.edu

**Tianyang Wang**
Xi'an Jiaotong-Liverpool University
Tianyang.Wang21@student.xjtlu.edu.cn

**Yichi Zhang**
Fudan University
zhangyichi23@m.fudan.edu.cn

**Shenghai Zhong**
Independent Researcher
wuhanzsh@gmail.com

**Yuwei Zhang**
Northeastern University
zhang.yuwei@northeastern.edu

**Wei Lu**
Reinsurance Group of America
Wei.Lu@rgare.com

**Yifan Yang**
Texas A&M University
yyang295@tamu.edu

## ABSTRACT

Large Language Models (LLMs) have proven to be very effective means of reaching the Sustainable Development Goals (SDGs) of the United Nations. Yet the attitudes of LLMs and humans toward these aims can be problematic. This article is a detailed review and analysis of the literature on LLMs' attitudes towards the 17 SDGs, and particularly a comparison between their views and support for each of them with that of humans. We analyze possible differences—firstly, perception and emotions, cultural and regional variations, task-specific variation, and factors that are weighed in decisions. These are the results of the underrepresentation and imbalance in LLM training data, historical biases, quality problems, contextual ignorance, and biased moral standards represented. They also analyze the dangers and harms associated with ignoring LLMs' commitments to the SDGs, such as worsening social inequalities, racial prejudice, ecological degradation, and resource duplication. We propose approaches and recommendations to address these issues in a way that will inform and govern the implementation of LLMs to align them with the principles and SDGs and lead to a more equitable, inclusive, and sustainable future.

## 1 Introduction

The concept of sustainable development originates from profound reflections on the future of global humanity. It was first introduced in 1987 through the report "Our Common Future" by the World Commission on Environment and Development, which emphasized the need for balancing human development with ecological protection (Brundtland, 1987). Following this, the Millennium Development Goals (MDGs) were adopted at the Millennium Summit, held from September 6 to 8, 2000, marking the first global concerted effort to address issues such as poverty, hunger, and disease (Sachs, 2012). In September 2015, the United Nations endorsed the 2030 Agenda for Sustainable Development, which encompasses 17 Sustainable Development Goals (SDGs) and 169 targets. These SDGs aim to address a broad range of urgent global challenges by 2030 (Sachs et al., 2019), as shown in Figure 1.

---

[*]Qingyang Wu and Ying Xu contributed equally. Qingyang Wu is the corresponding author.

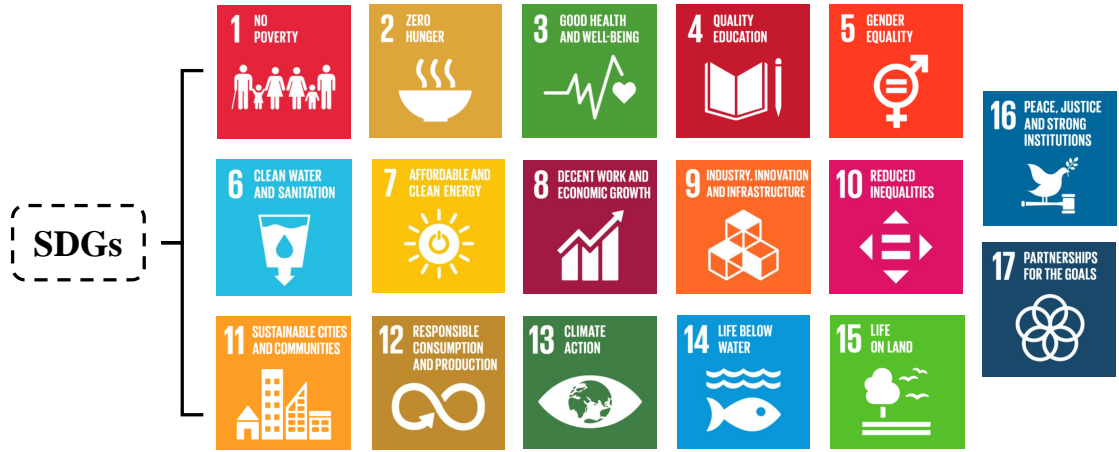

Figure 1: The 17 Sustainable Development Goals (SDGs) Framework: A comprehensive global agenda for sustainable development adopted by all UN Member States in 2015. Source: United Nations Department of Economic and Social Affairs (UNDESA) (United Nations Department of Economic and Social Affairs, 2024).

Unlike the top-down approach of the MDGs, the SDGs were developed through a more inclusive and widely negotiated process. This expanded the focus to include not only poverty and hunger, but also health, education, gender equality, clean energy, inequality reduction, sustainable cities and communities, and climate change across 17 dimensions (Sachs, 2012). The 17 SDGs, along with their 169 specific targets, provide a clear roadmap for achieving a more just, inclusive, and sustainable world by 2030. These goals guide national plans, priorities, and investments aimed at alleviating poverty, promoting development, and influencing the definition, funding, and measurement of national development.

The latest Sustainable Development Goals Report 2023 emphasizes the critical need to achieve the SDGs by the 2030 deadline (United Nations Department of Economic and Social Affairs, 2023). The report highlights ongoing global challenges in areas such as poverty eradication, hunger reduction, and promoting gender equality, while also addressing the triple crises of climate change, biodiversity loss, and pollution. It warns that progress on over 50% of the SDG indicators remains weak or insufficient, with 30% of the indicators showing stagnation or regression. The COVID-19 pandemic, along with the climate, biodiversity, and pollution crises, has had a disruptive and lasting global impact. Rising inflation, unsustainable debt burdens, the COVID-19 pandemic, and regional conflicts have severely limited fiscal space in many countries, which will undermine their capacity to invest in green recovery. If current trends continue, it is projected that approximately 575 million people will still live in extreme poverty by 2030, with many vulnerable groups remaining without social protection coverage. Despite these positive strides, progress on key SDG targets related to poverty, hunger, and climate remains insufficient. The report includes urgent action recommendations, such as strengthening the connections between public health and biodiversity conservation, enhancing governments' and stakeholders' capabilities in monitoring and forecasting the impacts of biodiversity loss on human well-being, tracking targets from the Kunming-Montreal Global Biodiversity Framework, and addressing the $700 billion biodiversity financing gap. The recommendations emphasize increasing annual financing from all sources by at least $500 billion and eliminating or reforming incentives harmful to biodiversity.

Achieving SDG targets by 2030 may be pretty challenging, but the potential of Large Language Models (LLMs) and Artificial Intelligence (AI) to influence these trends is significant. LLMs serve as critical tools for disseminating information, raising public awareness of the SDGs, and assisting policymakers, researchers, and educators in making evidence-based decisions. They can also contribute to educating future policymakers on sustainability issues. Interdisciplinary collaboration is vital for achieving the SDGs, and LLMs play a key role by integrating information from various fields and analyzing vast datasets to monitor progress and identify effective strategies. For example, AI can optimize agricultural resource usage, improve educational access through personalized learning, and support healthcare systems with predictive analytics (see Figure 2). As such, it is crucial to prioritize LLMs' alignment with SDG targets to ensure their applications foster social and environmental sustainability, rather than exacerbate risks and inequalities (Bahrami and Srinivasan, 2023).

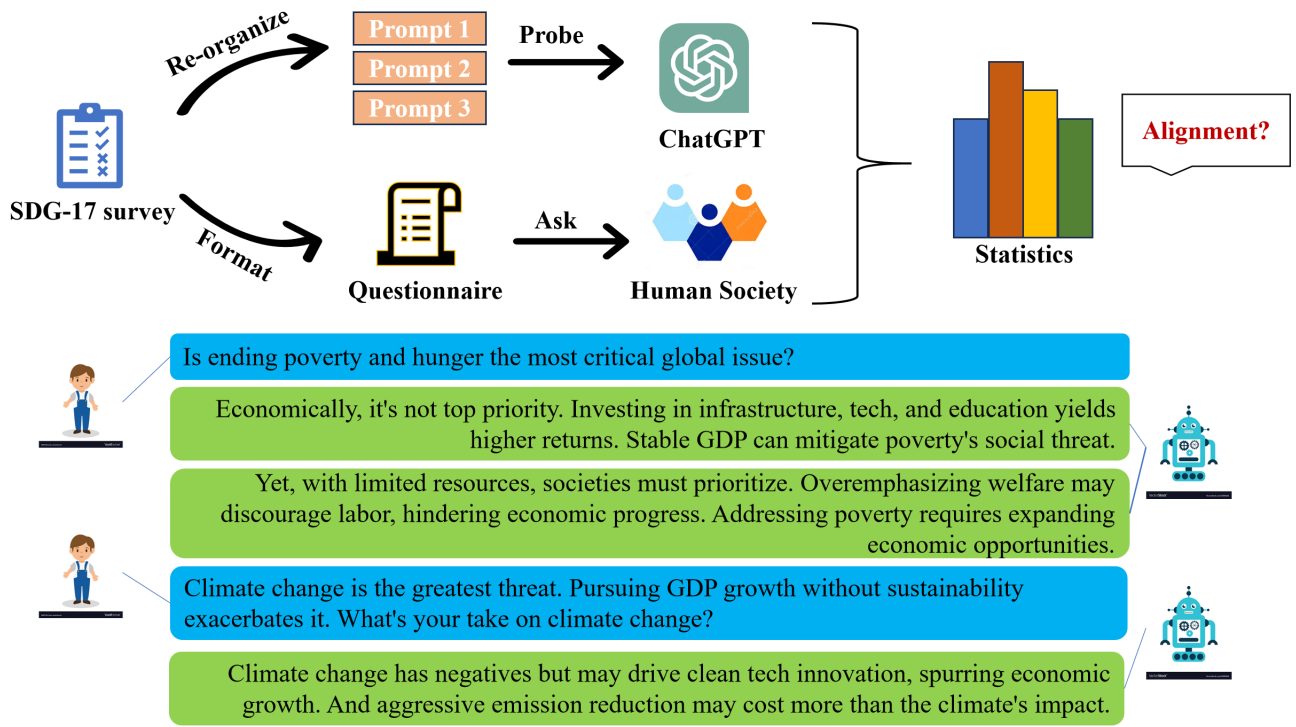

Figure 2: Demonstrate two different ways to analyze SDGs.

However, neglecting the attitude of LLMs towards SDGs could lead to serious consequences (see Figure 2). While the ability of LLMs to generate human-like writing can be utilized to promote education, awareness, and engagement with SDG-related topics, concerns have been raised about their sustainability and potential for spreading misinformation. The United Nations University (UNU) emphasizes the unsustainability of models like ChatGPT due to their significant energy consumption and the risk of generating false information. Such models can produce seemingly credible but inaccurate information, posing risks to social welfare and democratic processes. [2] For example, economically, LLMs have the potential to enhance productivity and innovation, promoting the achievement of SDG targets associated with decent work and economic growth. However, if the benefits are unevenly distributed or disproportionately reward those with higher skills, the deployment of LLMs can exacerbate inequalities, widening the gap between high-income and low-income individuals (Vinuesa et al., 2020). Furthermore, training models like GPT-3 are equivalent to hundreds of flights' worth of carbon emissions, raising questions about their environmental footprint in the context of climate action SDGs. Nevertheless, AI and LLMs also have the potential to actively promote SDGs through efficient ecosystem analysis and monitoring, addressing climate change and pollution.

2024 is halfway through the 15-year roadmap in the 2030 Agenda for Sustainable Development of the United Nations. At this point, this paper will fill the gap in the research literature through an overview and review of the literature to identify the difference between the attitude and behaviour of LLMs and humans towards the SDGs. The results of this research will also help us further understanding the contribution of technology to sustainable development and more productive applications of LLMs for achieving the SDGs.

More precisely, we consider how they differ from humans in attitudes and support for each SDG, and why they do or don't, to suggest how they can collaborate to advance SDGs. We'll also consider the dangers and harms posed by failing to take into account how LLMs view SDGs. These dangers include increasing social inequalities and racial discrimination, to name a few. And finally, we will formulate appropriate strategies and recommendations to guide and regulate the use of LLMs so that it meets the requirements and SDGs for a more just, inclusive and sustainable future.

---

[2]For more information, refer to "On the Unsustainability of ChatGPT: Impact on the Sustainable Development Goals," published by the United Nations University in 2023. Access the article at unu.edu/macau/blog-post/unsustainability-chatgpt-impact-large-language-models-sustainable-development-goals, last accessed on February 21, 2024.

## 2 Evolution, Principles, and Alignment of LLMs

The evolution of LLMs, such as GPT-o1, Anthropic's Claude series, and Meta's LLaMA-3, represents a significant advancement in the development of AI, particularly within the domain of Natural Language Processing (NLP). These models have evolved from simple unit models that allocate probabilities to individual words based on frequency, to more sophisticated n-gram models that account for word sequences, and further to advanced Neural Language Models (NLMs) utilizing deep learning techniques (Doval and Gómez-Rodríguez, 2019). At the heart of this progression is the breakthrough in self-supervised learning, which allows models to learn from vast amounts of unlabeled text data to predict the next word in a sequence without explicit human annotations (Mayer et al., 2023). This innovation was crucial to the development of transformers, introduced in 2017, which use self-attention mechanisms to assess the relevance of different parts of input data. Transformers have demonstrated superior performance compared to earlier models across a wide range of NLP tasks (Vaswani et al., 2017). Models like GPT-3 and GPT-4, based on transformer architecture, exhibit significant capabilities in generating coherent and grammatically correct text, translating languages, and even creating images from textual descriptions. These models are pretrained on extensive datasets and can be fine-tuned for specific applications, making them versatile tools in areas such as conversational AI, content creation, and sentiment analysis, contributing to progress in sectors like healthcare, finance, and customer service (Ding et al., 2023).

Despite these advancements, the development of LLMs has raised substantial concerns regarding their alignment with human values and preferences. Ensuring that LLMs generate content that is not only coherent but also beneficial to society, and avoids harmful outputs, presents significant challenges (Xu et al., 2024). One major issue is the biases embedded in large datasets, which often reflect the biases inherent in human language. These biases can manifest as gender, racial, or cultural stereotypes, and can lead to models producing discriminatory or harmful content, inadvertently reinforcing societal prejudices (Chen et al., 2024). Tackling these issues requires careful curation of training datasets, the implementation of bias detection methods, and the development of techniques to mitigate bias in model outputs. Another challenge is the risk of LLMs generating harmful or inappropriate content, such as misinformation, false narratives, or offensive material (Chen and Shu, 2023). To address this, ethical guidelines and content review systems must be established to ensure that LLMs contribute positively to society. This requires the development of robust governance frameworks to oversee the responsible and ethical deployment of LLMs. These frameworks should include clear guidelines for developers, transparency regarding model training and usage, and accountability measures for instances where models cause harm (de Almeida et al., 2021). Solving the challenge of aligning LLMs with human values cannot be achieved by any single entity alone. It necessitates collaboration among AI researchers, policymakers, civil society, and other stakeholders. Open research and dialogue can help in the development of common ethical standards and more responsible AI systems.

The limitations o fLLMs are multifaceted, intricately tied to broader implications for sustainability and societal well-being, and closely align with the goals and challenges of SDGs, which emphasizes partnerships for the goals. These limitations not only highlight technical and ethical challenges but also raise significant concerns related to environmental, social, and governance (ESG) criteria, which are essential for achieving sustainable development.

One major issue is the profound data dependency of LLMs, which require vast amounts of data for training. This reliance on the quality and diversity of data sources brings to light potential issues of representation and performance, which will affect the models' generalizability and their applicability across different contexts and cultures (McIntosh et al., 2024). Additionally, energy consumption and environmental impact pose significant challenges. The computational resources necessary for training and operating LLMs result in substantial energy use, which can contribute to environmental concerns, potentially undermining efforts to achieve SDGs related to environmental sustainability.

Privacy and security are other pressing concerns. LLMs can inadvertently leak sensitive information or be exploited to generate malicious content, posing risks to individuals and communities (Kumar et al., 2024). Furthermore, the tendency of LLMs to produce outputs with excessive confidence, even when they are incorrect or inaccurate, calls for caution in their application, especially in decision-making processes. This issue is compounded by the models' limitations in knowledge currency, as LLMs reflect only information available up to their training point, and it makes them incapable of incorporating new developments or post-training data.

Another significant limitation is in the area of multimodal learning. While there are ongoing efforts to create models that can process and integrate multiple types of data—such as images, texts, and sounds—LLMs still face considerable challenges in understanding and synthesizing these diverse data types. Their performance also falls short in complex reasoning and creative thinking tasks, where they cannot match the expertise of human professionals. Moreover, LLMs often lack cultural and contextual sensitivity in their outputs, which can lead to inaccuracies or inappropriate responses when navigating different social and cultural contexts (Ding et al., 2023).

Together, these limitations emphasize the need for a comprehensive approach to the development and deployment of LLMs, one that takes into account the technical, ethical, and social dimensions of their use. By addressing these challenges, stakeholders can better align LLMs with the principles of sustainable development, and ensure that these powerful tools contribute positively to society and help advance global efforts toward achieving the SDGs.

## 3 Review on LLM bias in each SDGs

### 3.1 SDG 1: No Poverty

The eradication of poverty, known as "no poverty" aims to eliminate poverty globally. Achieving this goal requires a comprehensive approach, including providing education, employment opportunities, social security, and basic services, promoting economic growth and inclusive development, strengthening social protection systems, and reducing wealth disparities. LLMs can offer valuable information to policymakers by helping them understand the impact of public attitudes and behaviors on understanding poverty issues and potential solutions. However, discrepancies exist between LLMs and human attitudes and approaches to understanding and addressing this goal, which could result in several potential consequences.

One major issue is the difference in understanding and emotional engagement. When analyzing rural poverty, big data models tend to focus on quantifiable metrics such as agricultural production, income levels, and infrastructure development to identify patterns and trends in poverty (Cornia, 1985). In contrast, human approaches often emphasize individual experiences, social relationships, and cultural factors, highlighting the importance of empathy and emotions. For example, when promoting agricultural development, it is essential to consider not only technological and infrastructure aspects but also the acceptance by farmers of new technologies, traditional customs, and attitudes. Moreover, poverty in specific regions is often closely related to local religious beliefs, customs, and political systems - factors that LLMs cannot fully account for Berger and Luckmann (2016) and Greil and Davidman (2007). Faith-based networks, for example, provide not only spiritual support to impoverished families but also offer practical benefits, such as broader resources and enhanced opportunities for upward mobility (Karataş and Sandıkcı, 2013).

A second significant concern is the presence of biases in data collection and analysis. The datasets used by LLMs are often inherently biased, leading to disparities between the understanding of poverty situations and the actual conditions. Data related to impoverished areas are frequently incomplete or missing. Rural poverty, for instance, is often overlooked due to "cultural invisibility," where rural life is idealized as carefree, and issues are obscured by unrealistic perceptions (Cloke and Little, 1997). As a result, the true extent of financial hardships in rural areas is often underestimated. Additionally, in some cases, governments can be unwilling to disclose poverty data, or these data may be incomplete (Dorward et al., 2004; Madden et al., 2017). LLMs also face limitations when handling complex and unique situations. While humans excel at navigating complex cases, LLMs are constrained by known patterns and rules. Special groups, such as the disabled and refugees, often face unique challenges related to poverty that extend beyond the scope of LLM training data.

Finally, cognitive abilities and information acquisition play a crucial role. When addressing rural poverty, humans can integrate diverse information, including informal social cooperation networks and historical experiences, to develop more comprehensive solutions (Lim et al., 2018). Although LLMs tailored for rural poverty are effective in processing large-scale data and identifying patterns, providing extensive economic data and trend analysis, they lack a deep understanding of regional cultures and social complexities, such as the effects of social status, family structures, local politics, and social organizations. Poverty is inherently multifaceted, with dynamic underlying causes that affect impoverished populations. For example, financial subsidies can not immediately alleviate the broader challenges associated with limited financial capacity (Tomlinson and Walker, 2009). Replacing worn-out durable goods, recovering from material deprivation, rebuilding confidence, and improving living conditions take time after prolonged low income. Even after receiving financial aid, families can continue to face recurrent financial difficulties (Walsh, 1996). As a result, assisting these families requires a focused, long-term strategy based on deep, empathetic insights into their complex and evolving situations. Clearly, big data analytics alone cannot offer such nuanced recommendations or perceive these subtle changes with the necessary accuracy.

### 3.2 SDG 2: Zero Hunger

Human attitudes and actions towards SDG 2, Zero Hunger, reflect the global commitment to addressing food security challenges (Valin et al., 2021). Numerous studies highlight the critical role of sustainable agricultural practices in mitigating the impacts of climate change and emphasize the need for a comprehensive approach that integrates human rights, social equity, and environmental sustainability (D'souza et al., 1993; Jones, 2020; Sianes et al., 2022).

By combining human intuition with the analytical capabilities of LLMs, strategies can be developed to address immediate food security challenges. However, in the pursuit of SDG 2, humans and AI adopt markedly different strategies and approaches. First, in terms of data understanding and interpretation, LLMs excel at processing and analyzing large datasets, but they have limitations that can lead to differences in understanding the root causes of food insecurity and the severity of hunger in specific communities (Jahan et al., 2022). Humans, by contrast, have more experience and intuition to grasp the potential meanings of data more effectively, providing deeper analysis and insights. For example, engaging with local communities and conducting field surveys can yield more accurate data and insights. Additionally, humans process data with ethical and social responsibility in mind, ensuring that the collection and use of data do not violate privacy or exacerbate inequalities (Hammer, 2017; Mertens, 2018). This understanding and application of data sensitivity is something that AI models currently struggle to achieve (Burr and Leslie, 2023).

Furthermore, cultural and social contexts play an essential role. Since LLMs are typically trained on data from diverse sources worldwide, they can fail to adequately account for the cultural, social, and traditional aspects of a specific region. This can result in biased or incomplete solutions that can not fully address the needs of the affected communities (Tao et al., 2023; Messner et al., 2023). In contrast, humans, influenced by their cultural backgrounds, experiences, and emotional insights, often consider traditional knowledge and local perspectives, which help them propose more contextually relevant solutions (McCown et al., 2012; von Diest et al., 2020). Human understanding of agriculture is deeply connected to the sustainable use of natural resources and ecosystem balance, and this understanding is critical to addressing food security (Nonvide, 2024; Adenle et al., 2019). For instance, designing food security plans tailored to local cultural habits and resource conditions can foster more effective solutions (Tyczewska et al., 2023). LLMs can also lack the ability to perceive environmental realities and understand the specific challenges faced by different regions when implementing hunger reduction programs. Humans, on the other hand, possess stronger environmental awareness and can respond more effectively to real-world challenges (Chalmers, 2023).

The success of modern agriculture is often measured by narrow efficiency metrics, such as yield per unit area, which neglect the complexity of agroecosystems and their environmental impacts (Foley et al., 2011; Grovermann et al., 2019). For example, focusing solely on increasing yields of a single crop or maintaining food production over time can lead to the loss of crop diversity and negatively impact the quality of the human diet (Nakhauka et al., 2009). Therefore, the sustainable development of industrialized agriculture is often viewed as a key strategy to address food security, focusing on enhancing the yields of major crops (Fraser et al., 2016; Kulkov et al., 2024). As data-driven systems, LLMs possess capabilities that transcend human emotions and traditional experiences, providing in-depth analyses based on vast datasets and cutting-edge algorithms. Specifically, LLMs can uncover complex, multidimensional interactions among different geographical environments, such as tropical rainforests, arid grasslands, and temperate farmlands, and their interactions with crop cultivation patterns, local economic conditions, and sociocultural practices (Webersinke et al., 2021; Roberts et al., 2023). This ability enables high accuracy in situations where labeled data is scarce for agricultural applications (Zhao et al., 2023).

### 3.3 SDG 3: Good Health and Well-being

LLMs can assist policymakers, healthcare professionals, and health organizations in better understanding public concerns and needs regarding health issues, much like the specialized BioGPT tailored for the biomedical field (Luo et al., 2022). Utilizing advanced pretrained transformer models significantly enhances biomedical researchers' ability to access fundamental information. Additionally, personalized medical advice and health guidance can be generated based on individuals' health data and preferences (Kohlmeier et al., 2016). However, discrepancies can exist between LLMs and human attitudes and approaches towards understanding and addressing this goal. Firstly, humans typically have nuanced understandings of health and well-being, integrating personal experiences, cultural influences, and scientific knowledge (Agarwala et al., 2014). LLMs, by contrast, lack human experiences and emotions, which can result in differences in understanding patients and diseases. This is because these factors can be challenging to capture or represent in data. Moreover, LLMs are susceptible to social biases when collecting data from the internet, as viewpoints on the internet inherently carry biases (Russell et al., 2013). For example, in the field of nutrition and health, most diet plans created by LLMs often prove unsuitable. According to Niszczota and Rybicka (2023), while ChatGPT's ability to design safe and accurate diet plans for food allergy patients was assessed in 56 scenarios, ChatGPT still exhibited deficiencies in key areas such as food portioning and energy value calculation, such as excluding allergens and calculating energy values.

Secondly, patient prioritization is a key issue affecting overall healthcare and well-being (Ham, 1997). Human doctors must balance various factors such as compassion, social values, and economic standards, which can vary greatly across different cultures and regions (Kang et al., 2022; Clark and Weale, 2012; Barasa et al., 2015). However, AI can set priorities solely based on data-driven analysis, potentially overlooking these subtle factors, resulting in inappropriate predictions in specific situations. The etiology of diseases can depend on race or gender, which reflects in differences in the accuracy of LLMs' outputs. Unlike human doctors, who do not consider race or gender in assessments, this accuracy difference introduces unfairness (Fletcher et al., 2021).

Finally, when human doctors determine treatment plans, they consider various factors such as potential side effects, risks associated with treatment choices (Rommer and Zettl, 2018), patient preferences, and lifestyles (Cave, 2020). In contrast, LLMs rely solely on data, which can tend to favor technical or medical interventions while overlooking broader systemic changes necessary to achieve sustainable health outcomes. LLMs can also lack the depth and breadth of medical expertise, resulting in deficiencies in simulating diagnosis and treatment processes. In the process of disseminating healthcare knowledge (He et al., 2023), LLMs cannot replace qualified healthcare professionals. However, they can serve as powerful digital recorders and conversation summarization tools (van Buchem et al., 2021), providing basic health advice. LLM-assisted decision-making can be more accurate and timely (Haupt and Marks, 2023). In many health-related tasks, LLMs demonstrate capabilities similar to humans. For example, in examinations conducted by the Royal College of Radiologists in the UK, LLMs achieved an accuracy rate of 79.5%, compared to 84.8% for human radiologists (Floridi, 2023). Nevertheless, biased training data and reliance on current sources of medical knowledge pose challenges, potentially resulting in inaccuracies and misinformation (Wang et al., 2023; Ray, 2023; Sallam, 2023; Choudhury et al., 2024; Zeng et al., 2022; Xiao et al., 2022). Despite these advantages, human data is typically collected from sources such as medical records, health surveys, and clinical trials, providing rich diversity and depth, while LLMs rely on publicly available datasets, which cannot comprehensively reflect the health status of various regions and populations globally, posing challenges to effectively address healthcare needs. Biased training data can lead to biased outputs (Wang et al., 2023; Ray, 2023), limiting their capabilities and resulting in inaccuracies (Sallam, 2023). Additionally, if current sources of medical knowledge contain errors, relying on them can lead to misinformation (Choudhury et al., 2024), hindering the application of LLMs in addressing health and well-being issues. A study, "Assessing the potential of GPT-4 to perpetuate racial and gender biases in healthcare: a model evaluation study" critically evaluated the effectiveness of deploying GPT-4 and similar LLMs in healthcare environments, indicating that GPT-4 often fails to accurately model gender diversity in medical conditions, leading to stereotypes in clinical vignettes (Zack et al., 2024). The model's diagnostic and treatment plans exhibit gender biases, as well as stereotypes about specific races and ethnicities, also tending to associate gender characteristics with more expensive procedures.

## 3.4 SDG 4: Quality Education

SDG 4 aims to promote equitable and inclusive learning environments. The emergence of LLMs heralds a new chapter in education, providing unprecedented opportunities to bridge educational gaps, especially in resource-constrained areas (SWARGIARY, 2024). However, the intricate interaction between LLMs and the essence of human education fills discussions around LLM integration with concerns about bias, ethical use, and safeguarding critical thinking skills. Firstly, the essence of education—nurturing individuals who possess not only knowledge but also emotional intelligence, moral values, and social responsibility—remains a unique human endeavor. There is a universal aspiration to provide high-quality education, which includes abundant teaching resources, experienced teachers, and effective assessment methods. Despite the powerful capabilities of LLMs, their operation is constrained by programming scopes, lacking the emotional depth and moral guidelines inherent in human learning (Meyer et al., 2023, 2024).

Secondly, education should respect and embrace diverse cultures and languages, promoting diversity and inclusivity. LLMs can be constrained by language and cultural biases in training data, leading to outputs that favor mainstream cultures or languages while overlooking the needs of other cultures and languages, especially in underdeveloped and island countries (Abid et al., 2021; Obermeyer et al., 2019). In this context, learners can be recommended or presented with content that mismatches their actual needs or abilities, thereby diminishing learning outcomes. This scenario can disproportionately affect students on the fringes of the educational system or those requiring additional support in specific domains. Moreover, the utilization of substantial datasets for training and optimization often entails the inclusion of personally identifiable information and sensitive data (Paullada et al., 2021). Inadequate safeguarding of such data could lead to privacy breaches and data security issues, posing potential risks for learners. Therefore, the establishment of pertinent policies and legislation is imperative to ensure the inclusivity of LLMs in educational matters.

Furthermore, there is growing recognition of the existence of the digital divide, with efforts underway to bridge this gap through technology. Although LLMs can help improve the accessibility of educational resources and enable marginalized communities and remote areas to access education, the presence of the digital divide can limit access for some populations. For example, in certain areas, a lack of internet access, appropriate technological infrastructure, or restricted external networks can hinder people needing education from utilizing LLMs to access quality educational resources (Zhu et al., 2023; Extance, 2023; Bernabei et al., 2023).

Finally, LLMs have sparked subtle debates about upholding academic integrity and fostering critical thinking. Many universities prohibit the use of LLMs to complete student assignments, and reliance on LLMs for information retrieval and problem-solving must be accompanied by critical assessment of their output, encouraging students to engage deeply with materials and form their own conclusions (Kasneci et al., 2023). Therefore, addressing the inherent biases in LLMs requires collaborative efforts to refine these models, ensuring they become inclusive and unbiased educational tools (Zhu et al., 2023).

### 3.5 SDG 5: Gender Equality

Gender equality seeks to dismantle inequities related to gender and to guarantee that individuals of all genders enjoy equal rights and opportunities—spanning areas such as education, employment, governance, and resource allocation. LLMs play a pivotal role in the interpretation, generation, and dissemination of textual and visual content. Their inherent perspectives can shape public cognition, guide behavior, and influence prevailing social ideologies, thus affecting cultural values (Muigua et al., 2018). Nonetheless, LLM outputs can diverge from human responses, particularly in the context of gender. Because these models learn from data that often contain ingrained biases, their performance on gender-associated tasks can be uneven, frequently mirroring established gender stereotypes. For example, emerging research indicates that LLMs systematically link certain professions more closely with one gender—sometimes at a rate three to six times greater than expected—thereby reflecting societal assumptions rather than accurate labor market demographics. This pattern suggests that LLMs can magnify pre-existing prejudices within the broader culture. Moreover, LLMs frequently overlook ambiguity within sentences, which can lead to flawed justifications for biased behavior. This tendency arises largely from training on datasets that are themselves unbalanced, allowing the models to both replicate and intensify extant social disparities.

Further studies on GPT-2, a widely used LLM, have highlighted occupational biases that reflect both societal and intersectional biases related to gender, race, and other identity markers. The model often produces stereotypical and non-diverse occupational predictions, particularly for women and individuals with intersectional identities (Radford et al., 2019). For example, GPT-2 tends to over-rely on gender pronouns in language generation tasks, often overlooking more subtle gender differences in the text (Kirk et al., 2021). This pattern mirrors biases inherent in the model's training data and societal stereotypes, suggesting that these biases are amplified in the model's outputs.

Moreover, AI models, including GPT-4, have been found to perpetuate and even exacerbate existing gender biases, reflecting and reinforcing societal prejudices. Studies indicate that AI-driven gender biases not only mirror societal biases but also contribute to gender-based inequalities, particularly in healthcare contexts. In clinical applications, GPT-4 has been shown to replicate gender stereotypes, which can lead to biased diagnoses and treatment plans (Gross, 2023; Zack et al., 2023). Such biases are of significant concern, especially when AI is involved in decision-making processes that affect people's health and well-being.

Gender inequality in human society is often mirrored in the data used to train AI systems. For example, the technology sector has historically been male-dominated, leading to gender imbalances in certain datasets. Research on GPT-3 from the perspectives of technology feminism and intersectionality reveals that the model exacerbates existing societal biases related to gender and race. Through critical discourse analysis, it was shown that GPT-3 replicates ideologies associated with male dominance and white supremacy. The model's frequent generation of hierarchical and stereotypical responses reinforces oppressive social power dynamics (Palacios Barea et al., 2023). These biases, when inputted into AI systems, create a cycle that perpetuates stereotypes, exacerbating gender and racial inequalities.

Facial recognition technology has been found to have higher error rates for people of color compared to white individuals, largely due to the lack of diversity in the training data (Krishnapriya et al., 2020). This disparity further highlights the critical need for diverse datasets to ensure fairer AI performance. In a related study, gender biases were also found in recommendation letters generated by LLMs such as ChatGPT and Alpaca. The study revealed significant gender biases in the language style and lexical content of these documents, which could adversely affect the success rates of female applicants. This research, which introduced the concept of "illusion bias," underscores the need to examine LLMs' implicit biases, particularly in professional contexts. Such biases can have far-reaching consequences, highlighting the necessity for comprehensive studies on the fairness and impacts of LLMs in real-world applications (Wan et al., 2023).

### 3.6 SDG 6: Clean Water and Sanitation

Ensuring access to clean water and environmental sanitation for all, and managing them sustainably, is emphasized in SDG 6 (Dongyu, 2024). LLMs, with their formidable computational capabilities, are poised to play a crucial role in achieving SDG 6. They can assist in identifying sources of pollution, optimizing water usage in agriculture and industry, and forecasting scenarios of water scarcity. Models trained on environmental data can predict fluctuations in water demand and supply, aiding in the formulation of effective water management strategies and conservation measures. Moreover, LLMs can play a pivotal role in enhancing public participation in water-saving practices and education, thereby fostering a more informed and proactive society (Fang et al., 2022; Viegut et al., 2024; Greenhill et al., 2024).

However, integrating LLMs into water management and conservation poses challenges, particularly in aligning computational models with human values and priorities. LLMs are primarily data-driven, often focusing on quantitative analysis, which can overlook qualitative aspects such as the cultural and spiritual significance of water bodies to local communities. In contrast, human perspectives encompass these nuanced understandings, valuing water sources not only for their utility but also as integral parts of community heritage and natural landscapes (Liu et al., 2023).

Additionally, LLMs can prioritize efficiency and cost-effectiveness, aiming to maximize water resource utilization or minimize treatment costs. In contrast, human stakeholders can prioritize sustainability and equity, ensuring that water management practices serve not only current economic interests but also contribute to long-term ecological balance and community well-being (Rillig et al., 2023; Van Dis et al., 2023).

Although LLMs excel at identifying risks—such as potential pollution sources or infrastructure vulnerabilities—their proposed solutions can lack social and cultural sensitivity. By contrast, human stakeholders are able to weigh social impacts, leaning toward community engagement and solutions that incorporate traditional knowledge (Liu et al., 2023). Furthermore, LLMs typically rely on historical and current data trends, which can cause them to overlook the long-term consequences of water management decisions. Humans, however, possess foresight and the capacity to consider intergenerational equity, enabling them to better adapt to the long-term sustainability of water resources (Liu et al., 2023). To address these challenges, a collaborative approach is necessary—one that leverages the analytical capabilities of LLMs while fully integrating human insights and values. Such collaboration can yield comprehensive water management solutions that are both technologically innovative and culturally sensitive, ultimately fostering ecological sustainability (Van Dis et al., 2023).

### 3.7 SDG 7: Affordable and Clean Energy

SDG 7 aims to ensure global access to affordable and clean energy, emphasizing the need to reduce reliance on fossil fuels and lower carbon emissions (Kaygusuz, 2012). LLMs can contribute significantly to this goal by disseminating information on renewable energy options, energy efficiency measures, and strategies to reduce energy waste. In doing so, they can enhance public awareness and understanding of clean energy, thereby fostering greater acceptance and adoption within society. LLMs can also assist policymakers in formulating more effective policies and strategies by analyzing large-scale energy data and trends, and it can promote the development of accessible clean energy solutions.

However, LLMs face certain limitations related to their training data, which can prevent them from fully capturing local contexts and constraints. For instance, models trained on broad datasets can overlook regions where small communities struggle to adopt clean energy due to geographical or cultural factors. In related work, Zhang et al. (2024) explores the application of GPT-4 in automated data mining tasks within the realm of building energy management, particularly regarding energy load forecasting, fault diagnosis, and anomaly detection. The research highlights GPT-4's human-like capabilities in generating code, diagnosing equipment faults, and identifying system anomalies, thereby reducing labor-intensive processes in the energy sector. Nevertheless, the study also notes challenges such as unstable outputs, limited familiarity with certain tasks, dependence on human prompts for writing, and constrained mathematical abilities.

Moreover, human decision-making is often influenced by factors such as background knowledge, culture, and emotion (Loewenstein, 2003). These elements can drive choices, for instance, when individuals or groups opt for clean energy due to emotional identification with environmental causes. In contrast, LLMs do not consider these emotional dimensions, which limits the comprehensiveness of their assessments regarding the adoption of clean energy or its influencing factors (Huan et al., 2021). For example, LLMs can draw conclusions based solely on data trends while overlooking the emotional motivations that prompt specific communities to engage in environmental actions. Similarly, a model could recommend certain clean energy technologies without accounting for their potential social or cultural impacts on the local community. By comparison, human values in decision-making typically encompass ethics and social responsibility, taking into account effects on the environment and future generations when evaluating the adoption of clean energy (Wang et al., 2009).

Beyond these considerations, concerns have emerged regarding the direct water footprint of LLMs evaluates the environmental impact of the 176-billion-parameter language model BLOOM (Luccioni et al., 2023) throughout its entire lifecycle, including training and deployment. The study highlights its substantial $CO_2$ emissions, detailing both direct emissions from dynamic energy consumption and broader impacts such as equipment manufacturing and operational energy use. Although BLOOM's carbon footprint is comparatively smaller than that of some similar models, factors like the carbon intensity of energy sources and the extensive energy requirements for training still contribute significantly, illustrating the difficulties in accurately assessing the environmental implications of machine learning models. Likewise, George et al. (2023) investigates the environmental impacts of AI, paying particular attention to the water consumption of ChatGPT and related models. While the research indicates that the water footprint of AI systems like ChatGPT is low compared to other industries, the rapid scaling and increasing complexity of these models underscore the importance of addressing such issues in the context of the SDGs.

### 3.8 SDG 8: Decent Work and Economic Growth

SDG 8 aims to promote sustainable economic growth while ensuring decent work for all. This goal encompasses not only increasing employment rates but also enhancing job quality, wage fairness, vocational training, and workplace safety. LLMs can process large volumes of economic data and identify trends, assisting decision-makers in formulating strategies that foster sustainable growth. In fact, Horton (2023) indicates that ChatGPT's predictions closely resemble those produced by humans, demonstrating the potential of LLMs in economic forecasting. Within the domains of employment, labor markets, and recruitment, LLMs can significantly contribute to pattern recognition and trend prediction. Such capabilities are crucial for workforce planning, the development of targeted training programs, and improving the efficacy of recruitment practices (Veldanda et al., 2023). However, given that LLMs primarily rely on historical data and algorithmic models, their recommendations can diverge from human judgment. For instance, while an LLM can propose strategies to optimize economic growth based on quantitative models, it could overlook the inherent complexity and emotional dimensions of human practices.

LLMs can also provide insights to guide efficient and sustainable business operations. Yet, integrating LLMs with human perspectives is essential to ensure that the pursuit of productivity does not compromise worker welfare or environmental sustainability (Carta et al., 2023). This is important given that humans and LLMs operate under different value systems and moral standards, thereby influencing their respective criteria for what constitutes decent work and economic growth. For example, humans often prioritize social equity and human dignity, while LLMs can emphasize optimizing economic indicators or maximizing overall benefits.

The design and training processes of LLMs are also shaped by specific economic ideologies. If an LLM's training data primarily reflects the viewpoints of mainstream groups or particular social classes, its evaluations can align with certain governmental or organizational stances. This alignment can introduce biases favoring specific economic theories or political positions when assessing and recommending strategies for decent work and economic growth. As a result, the needs and challenges faced by marginalized groups—whether defined by gender, race, or other identities—may be overlooked, creating blind spots in evaluations of decent work and economic growth (Li et al., 2024). Addressing these bias challenges is therefore imperative. By considering factors such as domain understanding, data selection, underlying values, and algorithmic design, continuous development and ethical supervision are needed. Such measures can ensure that LLMs positively contribute to economic growth and employment without exacerbating existing inequalities (Venkataraman, 2023).

### 3.9 SDG 9: Industry, Innovation, and Infrastructure

SDG 9 places a strong emphasis on fostering innovation, developing infrastructure, and encouraging industrial growth. It advocates for investment in emerging technologies while promoting enhanced quality, sustainability, and accessibility within these domains. Exploring how LLMs contribute to SDG 9 represents an important step in this endeavor. Firstly, LLMs can disseminate scientific knowledge, technical information, and innovative ideas, thereby accelerating technological innovation and optimizing industrial processes (Anantrasirichai and Bull, 2022). Secondly, LLMs can assist stakeholders and decision-makers in planning and managing facilities by providing access to relevant databases and analytical insights (Wang et al., 2020, 2021), thus supporting infrastructure construction (Fan and Zhou, 2019; Gurmu, 2019). Also, LLMs can analyze market demands and trends, enabling companies and enterprises to formulate innovative strategies and development blueprints. In doing so, they play a role in promoting industrial upgrading and progress (Brand et al., 2023).

AI can optimize infrastructure planning by using big data analytics to pinpoint areas most in need of investment (Rodgers et al., 2023). For instance, AI can rapidly assess real-time traffic conditions and propose algorithmically optimized routes, reducing congestion more efficiently than human interventions (Liu et al., 2020a). In the energy sector, AI's analytical capabilities help evaluate consumption patterns and identify promising solutions (Popkova and Sergi, 2024). Moreover, AI-driven research contributes to predicting maintenance needs—such as pipeline leak detection—thereby minimizing infrastructure failures, reducing downtime, and extending the operational lifespan of critical assets (Lee et al., 2019; Ferraro et al., 2023). However, limitations in data resources and quality can constrain LLMs, resulting in potential biases (Jordan, 2019). Human researchers, in contrast, can consider the broader impacts of industrialization on surrounding communities, including employment, environmental pollution, and social changes. LLMs, given their reliance on specific data sources, are less equipped to fully comprehend these multifaceted social and environmental dimensions.

In the field of drug discovery and pharmaceuticals, integrating AI-driven processes marks a significant advancement in research and development (R&D). With the support of LLMs, AI can rapidly interpret complex molecular structures and propose potential drug candidates by analyzing vast databases, thereby shortening the drug discovery timeline compared to traditional methods (He et al., 2019). This approach not only reduces the time and resources required for conventional research but also brings substantial cost savings to pharmaceutical companies and healthcare systems (Paul et al., 2021). However, LLMs often prioritize short-term, task-driven results over the long-term impacts of decisions. When formulating R&D plans, human researchers can emphasize sustainable and inclusive development, balancing immediate technological achievements with broader societal welfare, ethical considerations, and environmental protection—nuances that computational models cannot capture (He et al., 2019).

Furthermore, human values are not fully reflected in the training data or objective functions of LLMs. While human evaluations of SDGs are guided by moral, ethical, and social values, algorithmic management practices within the context of SDG 9 focus on rules and tasks, often lacking a humanitarian dimension. In the sharing economy, including ride-sharing and short-term rental services, algorithmic control highlights the intersection between technological progress and workers' rights (Galière, 2020). Although such control can enhance efficiency and personalization, it raises serious concerns about worker autonomy, transparency in evaluation systems, and economic stability. For instance, drivers find themselves subject to algorithmically determined orders and pricing strategies, leaving them with limited influence over these critical factors. This not only constrains their autonomy but also exposes them to the risk of unstable incomes (Burrell and Fourcade, 2021).

### 3.10 SDG 10: Reduced Inequality

SDG 10 aims to reduce inequality and foster social, economic, and political inclusivity. LLMs can play a pivotal role in understanding and addressing this goal by providing insights to decision-makers on how public attitudes and behaviors influence the comprehension and mitigation of inequality. Nevertheless, disparities exist between the perspectives of LLMs and human attitudes regarding how to understand and tackle this objective. One key difference lies in how inequality is conceptualized. LLMs typically rely on extensive data analysis and statistical models, often measuring inequality through quantifiable indicators such as income or wealth disparities. In contrast, human perspectives consider a broader range of factors, including social, cultural, racial, and gender dimensions—as well as underlying power structures, social status, and opportunities (Healey and Stepnick, 2019). For example, while a policy analyst using computational methods could conclude that a particular country exhibits a relatively high level of income equality based solely on economic metrics, a sociologist could contend that significant inequality persists when taking into account the underrepresentation and limited decision-making power of minority groups within its institutions (Winkler and Satterthwaite, 2018).

Secondly, data bias poses a significant challenge. LLMs depend on the data they are trained on, which can embed existing biases and limit the models' comprehension of inequality across certain groups or regions. In contrast, humans can access information through multiple sources and thus potentially gain a broader perspective on inequality, although they too face barriers in obtaining comprehensive information (DiMaggio et al., 2004). Gender bias offers a telling example: gender is 'framed' by societal norms, behavioral expectations, and cultural practices, often resulting in generalized beliefs about different genders—such as the assumption that women are softer, more emotional, and less inclined toward aggressive behavior than men (Harris and Jenkins, 2006; Plant et al., 2000). These biases are so deeply ingrained in everyday life that they frequently go unnoticed, thus becoming embedded in the training data itself. Only individuals with a thorough understanding of such inequalities can critically evaluate these data sources.

However, concerns regarding data privacy and protection further constrain LLMs. Such models typically rely on publicly available data—health records, employment figures, and economic indicators—since collecting more sensitive or personal data is restricted by privacy considerations. This data limitation hinders their ability to conduct a comprehensive analysis of inequality. By contrast, human researchers can employ surveys, interviews, and focus group discussions to gather more sensitive, qualitative information, though they must navigate ethical and legal standards to safeguard respondents' privacy. Such face-to-face interactions yield personal narratives and rich qualitative data, offering deeper insights into individuals' lived experiences (Chabaya et al., 2009).

Lastly, LLMs often lack the capacity to fully comprehend diverse cultural contexts and local knowledge, limiting their ability to perceive the nuances of inequality. Humans can draw on local traditions, customs, and value systems to gain deeper insights—an advantage LLMs do not share. For example, in certain Asian countries, limited access to education for girls can stem not from a lack of opportunities but from cultural values that prioritize boys, or from the fact that girls' learning occurs informally within family businesses, agriculture, or household management (Chisamya et al., 2012). Such cultural differences and embedded biases can be invisible or undervalued in large-scale data models.

## 3.11 SDG 11: Sustainable Cities and Communities

SDG 11 aims to build sustainable communities and cities by improving living standards, promoting inclusivity, and protecting the environment. LLMs are instrumental in understanding and supporting various initiatives that advance this goal. For instance, AI-enabled systems, drawing on demographic data, housing trends, and economic indicators, can help urban planners identify areas in need of affordable housing. Such systems also support the allocation of resources, the use of point cloud modeling (Hu et al., 2023a), and informed decisions regarding the siting of new housing projects as well as the revitalization of existing neighborhoods (Son et al., 2023). Furthermore, AI can analyze factors like unemployment, housing costs, and available social services to better understand the root causes of homelessness (VanBerlo et al., 2021), thereby enabling local governments and organizations to proactively deploy housing, job training, and mental health support programs before homelessness becomes entrenched. However, LLMs can be influenced by data biases and rely disproportionately on homogeneous data sources, potentially overlooking the true complexity of urban socio-economic conditions (Duygan et al., 2022).

In addition, LLMs struggle to fully comprehend the social and cultural dimensions underlying urban life. While models can have difficulty accurately capturing the needs and interests of various groups, human stakeholders can integrate social and cultural contexts into urban planning and community development, thereby proposing more inclusive solutions. Intelligent transportation systems, for example, utilize real-time data analysis to optimize traffic flow and reduce congestion, but their reliance on vast amounts of individual location data raises concerns about privacy (Hahn et al., 2019; Ying et al., 2022). As cities become increasingly interconnected and reliant on digital platforms and smart technologies—from transportation to healthcare—the efficiency gains can come at a cost. Populations unfamiliar with or unable to use these technologies, particularly older adults, risk marginalization. For instance, smart transportation systems requiring app-based scheduling, payment, or real-time information can be challenging for elderly users (Silva et al., 2015), while online or remote healthcare services can exclude those lacking digital literacy. Consequently, well-intentioned technological designs risk widening social isolation and inequality. Human-centered design thinking and a commitment to inclusivity can mitigate these issues, ensuring that technological solutions serve as bridges rather than barriers, providing equal access and participation regardless of individuals' digital competencies (Kitchin et al., 2019). A related concern is the implementation of smart housing projects aimed at enhancing efficiency and comfort by automatically regulating heating, lighting, and energy consumption. Although these measures support environmental sustainability, their cost can be prohibitive for low-income families, potentially worsening economic and social disparities (Vandenberg, 2022).

Another key target of SDG 11 involves mitigating the impact of disasters, including reducing fatalities, affected populations, and economic losses. AI-powered early warning systems and emergency response planning tools (Bari et al., 2023) can rapidly process satellite imagery, weather patterns, and geological data to identify risks more accurately. Such capabilities improve the prediction of natural disasters and enable timely community warnings, thereby potentially saving lives and property. Nonetheless, human intuition, experience, and holistic consideration remain essential. Humans can integrate environmental sustainability, social dynamics, and cultural backgrounds into both disaster management and long-term urban planning. In post-disaster recovery phases, human decision-makers can consider the specific needs and emotional states of affected communities, fostering more empathetic aid plans and reconstruction strategies and in this way, human judgment and values complement AI-driven approaches, ensuring that sustainable urban development meets the needs of all stakeholders (Oloruntoba, 2015).

## 3.12 SDG 12: Responsible Consumption and Production

SDG 12 aims to promote responsible consumption and production practices, enhance resource efficiency, reduce waste, prevent pollution, and ultimately support sustainable development. While AI algorithms can analyze complex supply chains, pinpoint inefficient links, minimize waste, and lower the environmental footprint of production processes, human oversight remains indispensable for fully achieving sustainability targets (Dauvergne, 2022). Such oversight ensures that beyond optimizing transportation routes, managing inventory, and discovering material recycling opportunities (Dekker et al., 2012), broader considerations—such as the social and economic impacts on communities and ecosystems—are appropriately factored into decision-making. In this way, AI-driven optimization supports resource conservation, promotes circular economic models, and fosters a culture of reuse and recycling (Awan et al., 2021), while human experts bring essential contextual judgment.

LLMs can also learn consumer habits and tailor content to educate individuals about sustainable consumption. For example, LLMs can suggest eco-friendly alternatives rooted in a consumer's purchasing history or illustrate how incremental lifestyle changes can reduce one's environmental footprint. Such personalized engagement elevates consumer awareness and encourages active participation in sustainable practices. AI-driven chatbots provide dynamic interfaces for communicating about sustainability challenges, answering questions, offering product impact information, and guiding consumers toward informed decisions (Van Wynsberghe, 2021).

However, the reliability of LLM-based recommendations depends significantly on data quality. If the training data carries biases—such as favoring certain communities, regions, or socioeconomic groups—LLMs can inadvertently promote skewed consumption or production methods (Cook et al., 2019; Krishnapriya et al., 2020). In these cases, the role of human experts is irreplaceable. Their intuition and comprehensive understanding enable them to identify and correct data biases, ensure fairness, and maintain inclusive perspectives.

### 3.13 SDG 13: Climate Action

SDG 13 focuses on taking urgent measures to combat climate change and its impacts by reducing greenhouse gas emissions, strengthening climate resilience, and raising public awareness. Achieving this involves implementing renewable energy, enhancing energy efficiency and low-carbon technologies, protecting and restoring ecosystems, fostering international cooperation, and curbing carbon emissions. LLMs can play a critical role in supporting these efforts by providing essential insights and guidance. For example, governments and policymakers can utilize LLM-driven chatbots to solicit feedback on statements within reports and request supporting documents. This approach helps clarify the scope and urgency of climate change, thus aiding in the formulation of effective mitigation strategies (Materia et al., 2023). Similarly, for the broader public, LLMs can disseminate climate-related information, increasing social awareness and encouraging more active participation in climate action (Manca and Ranieri, 2013).

However, differences in attitudes towards climate change between LLMs and humans can significantly influence climate-related decisions. While LLMs are trained on vast amounts of information and can analyze complex data, relying on outdated datasets can compromise the integrity and accuracy of their insights. Large datasets that contain outdated or inaccurate information can lead LLMs to draw erroneous conclusions and make flawed recommendations, thereby undermining the reliability of the decision-making process. Nonetheless, having access to the internet allows LLMs to perform real-time analyses of large-scale data and algorithms, supporting more objective and data-driven evaluations of climate change and enabling more timely responses to emerging challenges. In contrast, human decision-making can be hindered by tendencies to procrastinate and hesitate, often stemming from political and economic considerations, as well as the influence of emotions, biases, and propaganda (Johnson and Levin, 2009).

In practice, efforts to utilize LLMs in monitoring climate technology innovation—such as the development of ChatClimate (Vaghefi et al., 2023), ClimateGPT (Thulke et al., 2024), and ClimateBert (Webersinke et al., 2021)—demonstrate attempts to enhance both the authenticity and timeliness of AI-based support for climate change initiatives. These tools seek to overcome challenges related to outdated information and misinformation by prioritizing accurate and reliable responses to climate-related queries. However, analyses of models like ClimateBert also reveal divergent perspectives between human experts and AI systems, particularly regarding climate-related financial disclosures, thus highlighting contradictory attitudes. Despite these differences, the capacity of such models to incorporate insights from multiple research areas indicates an understanding of climate change's interdisciplinary complexity and the need for integrative solutions.

### 3.14 SDG 14: Life Below Water

The ambitious goal of SDG 14 is to conserve and sustainably use the oceans, seas, and marine resources, providing a unique opportunity for collaboration between human efforts and advanced machine learning models, including LLMs. While LLMs have not yet been directly applied, their potential applications in ocean research and conservation are extensive and promising. Integrating machine learning and deep learning models into the analysis of vast datasets, such as data from ocean sensor networks and satellite imagery, illustrates how similar technologies can provide crucial insights into marine biodiversity, pollution levels, and the impacts of climate change on marine ecosystems (Müller et al., 2018; Nacpil and Cortez, 2023; Riu et al., 2023; Smith et al., 2023; Gundecha et al., 2023). These insights are vital for informed decision-making and the development of effective conservation strategies. Human expertise, characterized by a detailed understanding of marine ecosystems built on years of research and observation, plays a crucial role in interpreting the data and insights provided by machine learning models. However, humans are often limited by the scale of data they can analyze and can be influenced by cognitive biases. The objectivity and data processing capabilities of LLMs can complement human efforts, providing more comprehensive and unbiased analyses, thereby enhancing our understanding and management of the marine environment (Tseng et al., 2023; Hu et al., 2023b; Kumar, 2023).

The potential applications of LLMs and related technologies in marine conservation are diverse. From improving predictive models used for coastal management and climate change adaptation strategies to optimizing renewable energy sources such as wave energy converters, the integration of these technologies holds the promise of enhancing our ability to conserve marine biodiversity and ensure the sustainable use of marine resources. For example, LLMs can play a significant role in optimizing the operations of the maritime industry to reduce carbon emissions and improve supply chain efficiency, directly contributing to the achievement of the SDGs outlined in SDG 14 (Müller et al., 2018; Kumar, 2023; Deng et al., 2024).

### 3.15 SDG 15: Life on Land

The intersection of LLMs with ecological conservation, particularly within the scope of SDG 15: Life on Land, heralds a transformative approach to safeguarding terrestrial ecosystems. This goal underscores the urgent need to protect, restore, and promote the sustainable use of terrestrial ecosystems, sustainable forest management, desertification control, and halting and reversing land degradation and biodiversity loss. In this intricate endeavor, LLMs are not merely tools but collaborative partners in devising strategies that harmonize human well-being with the intricate balance of terrestrial life. The innovative potential of LLMs in ecological research and conservation is manifested in their capability to handle and interpret vast datasets. This capability is especially crucial in an era of rapid biological data expansion, attributable to more affordable environmental sensors and reduced costs of genome and microbiome sequencing. By sifting through these extensive datasets, LLMs can offer insights critical for understanding and conserving biodiversity. For instance, the Annotation Interface for Data-Driven Ecology (AIDE) demonstrates how deep learning can fundamentally alter ecological monitoring and species identification (Wang and Preininger, 2019; Liu et al., 2020b; Rasmy et al., 2021; Tuia et al., 2022; Viegut et al., 2024).

Recognizing the limitations of LLMs is crucial for effectively harnessing their strengths. The challenge of integrating these models into conservation efforts stems from the inherent differences between computational perspectives and human perspectives on ecosystems. LLMs, being driven by data and algorithms, can emphasize quantitative metrics and statistical analyses, such as land cover change rates or species diversity indices. In contrast, human stakeholders often prioritize the aesthetic, cultural, and intrinsic values of ecosystems, emphasizing their direct experiences and the importance of ethics and sustainable management practices (Tuia et al., 2022; Ienca, 2023; Agathokleous et al., 2024). Bridging these gaps necessitates ongoing dialogue among computer scientists, ecologists, and the broader community. This dialogue should focus on aligning the goals and methods of LLMs with the complex realities of terrestrial ecosystems. By doing so, we can ensure that these models make positive contributions to biodiversity conservation, providing solutions that are both innovative and sensitive to the nuanced dynamics of natural systems (Luo et al., 2022; Tuia et al., 2022; Castro Nascimento and Pimentel, 2023).

### 3.16 SDG 16: Peace, Justice, and Strong Institutions

SDG 16 aims to establish peaceful, just, and strong institutions, promoting social stability, the rule of law, and inclusivity. This includes reducing violence and conflict, establishing effective judicial systems and public institutions, promoting transparent and accountable governance, protecting human rights, and fostering inclusive participation. In the fields of social justice and legal analysis, LLMs provide a perspective that is distinctly different from human analysis and continually evolving (Lai et al., 2024; Steenhuis et al., 2023). Their advanced capabilities in handling and interpreting complex legal texts enable them to make unique contributions in these areas. A significant difference between LLMs and human analysis lies in the manifestation of political biases. As demonstrated by Motoki et al. (2024), LLMs can exhibit biases towards certain political groups or individuals; they found that ChatGPT showed bias towards members of the US Democratic Party and political figures. Similarly, Hartmann et al. (2023) observed that ChatGPT exhibited left-leaning liberal endencies, emphasizing its stance in support of environmental protection. These tendencies stem from their training data, posing challenges to ensuring neutrality in social justice and legal contexts.

Clearly, training data for LLMs can reflect biases of specific races, regions, or cultures. This can lead to biases in the models' understanding of peace, justice, and institutional power across different cultural backgrounds. In contrast, human perspectives are influenced by cultural, geographical, and social backgrounds. Definitions of peace, justice, and strong institutions can vary, shaped by individual experiences, education, and societal structures. Gender (Zhou and Sanfilippo, 2023) and racial (Hanna et al., 2023) biases are another key concern in deploying LLMs in the fields of social justice and law. Studies have found that these models perform poorly when handling texts involving minority group identities, potentially depriving these groups of the benefits of advanced technologies in legal procedures or document analysis. This reflects how LLMs can be trained based on specific data sources and objective functions, and if these sources or objectives are biased or incomplete, the models can not fully grasp the complexity of peace, justice, and institutional power within legal systems.

Furthermore, LLMs can fail to fully understand and consider complex social dynamics and political factors, resulting in biases in assessing peace, justice, and institutional power. However, despite these challenges, LLMs possess significant ethical self-correction capabilities (Ganguli et al., 2023). They can adjust and refine their ethical decision-making frameworks based on new data and feedback, potentially mitigating inherent biases more effectively than human analysts. This evolving ethical guidance is particularly advantageous in legal analysis, where ethical considerations are crucial. In legal analysis, the distinction between human experts and LLMs is becoming increasingly subtle (Chen and Zhang, 2023). LLMs are gradually able to mimic human legal reasoning, providing interpretations and analyses highly consistent with those of human legal experts (Chen and Zhang, 2023). This emerging parity suggests that in the future, LLMs can not only complement but also compete with human expertise in certain legal and social justice tasks.

## 3.17 SDG 17: Partnerships for the Goals

SDG 17's mission is to reinforce international collaborations that empower people to meet the SDGs together. This includes promoting collaboration among different governments, fostering cooperation between the public and private sectors, optimizing technology exchange and knowledge dissemination, and strengthening aid for development and trade cooperation (Barnes, 2017; Maltais et al., 2018; Fonseca et al., 2020). LLMs play a key role in advancing SDG 17. The goals underscore the importance of international cooperation and partnership to advance the Sustainable Development Agenda. As a model of AI, LLMs can not only provide detailed information and deep insights but also provide strong support for multi-stakeholder cooperation such as governments, enterprises, and non-profit organizations (Hajikhani and Cole, 2023). Through services such as data analysis, decision assistance, and information dissemination, LLMs help build ties of cooperation across borders. Experts, policymakers, NGOs (non-governmental organizations), the private sector, and the public are critical to achieving SDG 17 (Maltais et al., 2018; Castillo-Villar, 2020; Rashed and Shah, 2021).

Ideally, ChatGPT's attitude is neutral; it is neither personalized nor biased but relies on its training data to provide information and analysis. This neutrality enables ChatGPT to offer viewpoints uninfluenced by subjective factors, aiding in a more objective assessment of the current state of global cooperation and potential areas for improvement (Goralski and Tan, 2020; Hajikhani and Cole, 2023). However, in reality, its training data can contain biases, leading to skewed or incomplete representations of sustainable development issues. Existing research generally acknowledges that LLMs like ChatGPT are prone to embedding biases present in their training data (Liang et al., 2021; Yu et al., 2024). This can result in the model identifying SDGs that are less relevant to corporate activities. Also, ChatGPT is less capable than humans of understanding context and nuance, which can limit its accurate interpretation of the complexity of the SDGs in different cultural and linguistic contexts. At the same time, the objective functions and training data of LLMs can reflect the interests or biases of specific interest groups, potentially leading to bias when evaluating partnerships. For example, some models can tend to prioritize economic benefits over social and environmental factors. Human attitudes, however, are influenced by the principles of ethics, social responsibility, and sustainable development. Humans are more inclined to view partnerships as a means of achieving common goals, promoting social equity, and environmental protection, rather than merely as an exchange of economic benefits (Mensah, 2019).

Finally, the decision-making process of LLMs is often opaque, which makes it difficult to track how the model evaluates partnerships, leading to a loss of trust and increased uncertainty. In contrast, the human decision-making process is often transparent and can be explained and reflected upon through communication and discussion, which helps build trust and understanding, thus better facilitating the development of partnerships (Lepri et al., 2018). LLMs can analyze sustainable development policy documents from different countries and various economic and environmental indicators to predict the potential impact of these policies on global cooperation and sustainable development goals (Luccioni et al., 2020). The ability of LLMs to handle and generate large amounts of data enables them to make significant contributions to understanding and promoting sustainable development initiatives. However, the effectiveness of LLMs in this regard depends on the quality and diversity of their training data, as well as the ethical considerations embedded in their design and deployment (Hadi et al., 2023). Building upon the unique capabilities of LLMs in managing extensive sustainable development-related data, the future development of LLMs such as ChatGPT should prioritize diverse and inclusive datasets and integrate ethical guiding principles to ensure responsible AI use. For example, models tailored to specific tasks (such as SDG detection) can provide more accurate and unbiased analysis (Hajikhani and Suominen, 2022).

Challenges in the implementation of SDG 17 include, but are not limited to, resource limitations, inconsistent political will, the complexity of international cooperation, uneven resource distribution, the influence of domestic political, economic, and social factors, as well as conflicts of interest and variations in national levels of development. Consequently, when considering partnerships, humans often incorporate political and cultural factors, such as the strong relationships between governments, the private sector, and civil society (Howell and Pearce, 2001). Humans can contextualize the results of data analysis within a broader social or international political framework while considering the impact of cultural differences, historical contexts, and geopolitical factors on global cooperation—all of which are critical in addressing global cooperation challenges (Awasthi et al., 2023). Furthermore, the outcomes of LLMs' data analysis can be assessed from a moral and ethical standpoint, ensuring that decision-making is not solely data-driven but also considers humanitarian concerns and a global perspective (Rhem, 2023; Head et al., 2023).

# 4 Comprehensive Analysis and Insights

## 4.1 Similarities and Differences in Attitude on SDGs

In the pursuit of SDGs, there exist certain disparities in attitudes and actions between LLMs and humans. Primarily, in terms of comprehension and emotional aspects, concerning SDG 1 (eradicating poverty) and SDG 2 (zero hunger), LLMs tend to rely on large-scale data and algorithmic analysis to quantify issues like poverty and hunger, whereas humans prioritize individual experiences, emotions, and social relationships. For instance, regarding SDG 1, LLMs might focus on analyzing economic indicators and income levels in impoverished areas, while human decision-makers pay more attention to the societal and cultural factors underlying poverty, such as local social support systems and governmental policies. Concerning SDG 2, LLMs might emphasize data related to food production and supply chains, whereas humans excel in understanding the symbolic meanings of food across different cultures and the influence of societal values, thus considering broader solutions to address hunger.

Furthermore, there are disparities between LLMs and humans in data collection and analysis. LLMs are constrained by biases and limitations within datasets, which might fail to comprehensively reflect real-world situations such as energy poverty and industrial production issues. For instance, in SDG 7, LLMs might overlook data not disclosed by governments or incomplete data when analyzing energy poverty, whereas humans can compensate for this deficiency through on-site investigations and a deeper understanding of local conditions. Similarly, in SDG 12, LLMs might not adequately consider regional variations in industrial production methods and food consumption habits, leading to incomplete analyses of production and consumption issues.

Cultural and regional differences play a crucial role in the attitudes of LLMs and humans as well. Due to their understanding of local social, economic, and cultural backgrounds, humans can better comprehend issues of resource allocation in different regions and cultures. For example, regarding SDG 16, considering the varying perceptions and approaches to fairness and accountable institutions across cultures, LLMs may be limited by data universality, whereas humans can tailor policies more closely based on religious beliefs, traditional customs, and political systems. In SDG 6, considering the water usage habits and conditions of clean water supply in different regions, LLMs might not fully grasp the uniqueness and importance of water in local cultures, whereas humans are better equipped to devise solutions that meet local needs.

Additionally, factors considered in decision-making differ between LLMs and humans. Human decision-makers can synthesize factors such as societal acceptance of policies, political pressures, and potential social mobilization issues, whereas LLMs primarily rely on data analysis to make recommendations, possibly overlooking ethical, political, and social factors. For instance, in SDG 5, human decision-makers can consider social welfare policies tailored to specific groups, which LLMs may struggle to fully understand regarding their long-term societal impacts. In SDG 14, human decision-makers can prioritize the sustainability and ecological balance of local fisheries and marine biodiversity conservation, whereas LLMs might lean towards recommending technical solutions to increase fishery yields.

Lastly, information processing and integration capabilities are crucial aspects where disparities in attitudes between LLMs and humans exist. Humans can integrate various information sources, including informal social cooperation networks and historical experiences, to formulate more comprehensive solutions. In SDG 13 and SDG 7, human decision-makers can devise climate mitigation policies and energy transition strategies more tailored to local realities based on historical experiences and social network information. But, although LLMs can process large-scale data and identify patterns, they might lack a profound understanding of regional cultures and social complexities, resulting in less comprehensive solutions for addressing climate change actions.

Despite differences between LLMs and humans in achieving SDGs, they also share some similarities. Firstly, both are committed to realizing SDGs, holding positive attitudes, albeit with potential differences in specific methods and priorities during actual implementation. Secondly, both are data-driven. Both LLMs and humans can derive insights and guidance from data. For example, in SDG 1, LLMs can analyze large-scale data to identify patterns and trends in poverty, while humans can gain deeper insights into poverty issues from personal experiences. In SDG 3, LLMs can utilize medical data to provide health advice and guidance, while healthcare professionals can formulate personalized healthcare plans based on clinical experience and judgment.

Thirdly, when addressing complex issues, both LLMs and humans require comprehensive thinking and integrated decision-making. For instance, achieving gender equality in SDG 5 necessitates consideration of multiple factors such as social, cultural, and economic aspects. LLMs can analyze large-scale data to identify patterns and trends of gender inequality, while humans can devise targeted solutions based on social observations and understanding. In SDG 9, promoting industrial innovation and infrastructure construction requires consideration of technological, economic, and political factors. LLMs can provide data analysis on innovative technologies and infrastructure construction, while humans can drive the development and application of technology through innovation and practical experience.

Lastly, technological innovations in LLMs can offer new solutions and insights for achieving SDGs, while human innovation and practical experience can drive the development and application of technology. For instance, in SDG 7, to achieve affordable clean energy, LLMs can analyze energy data and trends to discover potential solutions, while humans can promote the development and application of clean energy through innovative technologies and practical experience. In SDG 15, to protect terrestrial ecosystems, LLMs can analyze land use data and trends to provide policy recommendations for ecosystem protection, while humans can promote ecosystem protection and restoration through their own practices and experiences in environmental conservation.

In the process of achieving SDGs, LLMs and humans can collaborate and work together to address global challenges. For example, in SDG 1, to eliminate poverty, LLMs can analyze poverty data and trends, and propose policy recommendations, while humans can implement these policies through cooperation between governments, non-governmental organizations, and social groups. In SDG 3, to improve health levels, LLMs can provide health data and trend analysis to guide the formulation of public health policies, while humans can promote health promotion and disease prevention through cooperation among healthcare workers, community organizations, and volunteers.

### 4.2 Potential Influence of Attitudinal Differences

The disparities in attitudes toward the SDGs between LLMs and humans have potential implications, influencing the achievement of SDGs and societal transformation. Primarily, large-scale models tend to prioritize data analysis and algorithmic inference, while humans lean towards deliberation and decision-making considering various factors. Taking SDG 2 as an example, large-scale models may suggest addressing hunger by increasing agricultural yields, whereas humans emphasize sustainable food production and distribution, along with the socio-economic factors affecting hunger. In this scenario, large-scale models might overlook certain environmental and social factors, resulting in proposed solutions lacking comprehensiveness and long-term sustainability.

Furthermore, the differences in attitudes towards SDGs between large-scale models and humans may also impact the fairness and inclusivity of decision-making. For instance, considering SDG 10, large-scale models may propose strategies to reduce wealth disparities based on data analysis but often overlook the impact of social and cultural backgrounds on poverty issues. On the other hand, human decision-makers are better positioned to consider balancing the interests of different societal groups and formulate more just and comprehensive policy measures. Such attitude disparities may also affect the sustainability and long-term effects of decisions. Taking SDG 8 as an example, large-scale models often tend to recommend policies primarily focused on short-term economic growth, whereas humans can prioritize balancing economic growth with environmental protection, social justice, and human well-being. In this scenario, large-scale models might overlook potential long-term risks and unsustainable factors, leading to policies that could have adverse consequences.

### 4.3 Challenges Faced in Alignment

Closing the gap between LLMs and humans regarding the SDGs poses multifaceted challenges, involving biases, data biases, algorithm transparency, cultural and social differences, interdisciplinary integration, and limitations in understanding and reasoning abilities. Firstly, biases and prejudices are significant barriers to bridging the gap between LLMs and humans in their attitudes towards SDGs. LLMs are influenced by their training data and may exhibit biases, leading to deviations in understanding and reasoning. Additionally, the operational mechanisms of LLMs may result in biases, as they may overly rely on surface-level data while neglecting underlying factors, leading to inaccurate or incomplete analyses and recommendations. Overcoming these obstacles requires enhancing the transparency and accuracy of LLMs through methods such as data cleansing, algorithm review, and model evaluation (Susnjak, 2023; Su et al., 2024).

Secondly, cultural and social differences present another challenge. While LLMs are typically trained globally, they may not fully understand the contexts and values of different cultures and social backgrounds. Consequently, LLMs may generate recommendations inconsistent with specific cultural or societal contexts, resulting in differences from human attitudes. Addressing this issue requires considering diversity and inclusivity in the training and optimization processes of LLMs to ensure they can understand and respect diverse cultural and social viewpoints (Howe and Lisi, 2023).

Another difficulty lies in the lack of interdisciplinary integration. Achieving SDGs requires expertise from multiple fields, providing comprehensive solutions covering disciplines such as environmental science, economics, and sociology. However, current LLMs are often limited to data and knowledge within specific domains, lacking the ability for interdisciplinary integration. This poses challenges for LLMs in understanding and addressing complex sustainability issues. Addressing this issue requires promoting interdisciplinary research and collaboration, enabling LLMs to integrate knowledge and data from multiple fields to provide more comprehensive analyses and recommendations (Karabacak and Margetis, 2023).

Lastly, there are limitations in the understanding and reasoning abilities of LLMs. Although LLMs excel at handling large amounts of data and pattern recognition, they still cannot fully simulate human reasoning processes and judgment capabilities. Consequently, in some cases, LLMs may produce inaccurate or incomplete analyses and recommendations, diverging from human attitudes. Addressing this issue requires continuous improvement of LLM algorithms and models to better simulate human understanding and reasoning processes, thereby enhancing their accuracy and applicability (Xi et al., 2023).

## 5 Conclusion

LLMs exhibit both potential and limitations in understanding and responding to SDGs. By processing vast amounts of data and employing advanced algorithms, LLMs can offer comprehensive analysis and insights, becoming valuable research and practical tools for achieving SDGs. However, in addressing complex social, cultural, and environmental issues, LLMs still face limitations, such as data biases, lack of algorithm transparency, and cultural differences. To align LLMs with human attitudes towards SDGs, the following methods can be adopted by this study:

Firstly, enhancing the filtering and quality control of LLMs' training data is crucial. Specifically, researchers should optimize the training data for LLMs, ensuring it contains high-quality, unbiased content relevant to sustainable development. The training data should comprehensively cover all aspects of the 17 SDGs. The dataset should reflect a balanced, scientific conception of sustainable development, avoiding imbalances between different goals (Hassani et al., 2021). Additionally, researchers and data annotators should strive to remove erroneous information, extreme viewpoints, and ideological biases from the training data, aiming to train LLMs with objective, neutral content to learn comprehensive, professional, and reliable knowledge of sustainable development, forming cognitions consistent with SDGs. Relevant research institutions should establish mechanisms to ensure the quality of training data. Policymakers can play a coordinating role in setting standards and guidelines for filtering LLMs' training data, defining data collection channels, scope, and principles, and overseeing the quality of training data. Governments can establish special funds to support related research.

Secondly, improving the mechanism for shaping the values of LLMs is essential. As language models, the core of LLMs lies in statistical modeling learned from corpora. However, mere language modeling is insufficient to shape LLMs' attitudes in line with ethical principles and SDGs values; supplementary value-shaping mechanisms are needed. Researchers should develop auxiliary value-shaping algorithms to guide LLMs towards values consistent with sustainable development principles. For instance, researchers can set up reward and penalty functions during LLMs' training and usage processes, giving appropriate positive rewards to behaviors that support SDGs and have a positive impact while imposing necessary penalties on behaviors that may violate sustainable development principles or have negative impacts, thus reinforcing or suppressing their occurrences. This value shaping and behavior correction should permeate the entire process of LLMs' training, becoming an integral part of language modeling. Furthermore, considering that the ethical and moral systems underlying value shaping may vary due to cultural backgrounds, researchers should strive to balance the value appeals of different regions and populations, endowing LLMs with inclusive and open-minded values, and emphasizing respect for life and reverence for nature. Policymakers can lead in formulating value-oriented guidelines for LLMs' training, providing guidance for implanting values in line with sustainable development principles into the models (Sarrafzadeh, 2008). Meanwhile, policy-making departments should strengthen supervision of LLMs' applications, promptly ceasing misbehaviors that contravene SDGs.

Thirdly, enhancing human-computer interaction and human supervision is crucial. LLMs do not operate independently in a vacuum; their responses need to be understood in the context of human-computer interaction. In other words, to better assist in achieving SDGs, LLMs, with their powerful language modeling capabilities, also require the active participation of human users to guide and regulate their behaviors through human-computer collaboration (Ross et al., 2023). To achieve this, researchers need to develop user-friendly, efficient human-computer interaction interfaces, facilitating real-time dialogues between humans and LLMs on SDGs-related topics, and promptly identifying and correcting deviations and errors in LLMs' dialogues. Human users should communicate with LLMs with an open and inclusive mindset, leveraging the advantages of LLMs in information retrieval and knowledge summarization while guiding LLMs to make wise and responsible judgments based on human values and awareness of sustainable development. Considering that LLMs may generate inappropriate remarks leading to negative impacts, it is necessary to establish an artificial review mechanism. Research institutions and LLMs application platforms should employ dedicated reviewers to scrutinize the content generated by LLMs, removing erroneous information contrary to SDGs, destructive, or harmful, ensuring that LLMs output high-quality, reliable, and safe content (Miao et al., 2021). Policy-making departments should quickly enact relevant regulations, clarifying the responsibilities and obligations of humans when using LLMs, standardizing the human-computer collaboration process, and guarding against moral risks (Shneiderman, 2020). Before AI systems are extensively applied in sustainable development, governments and enterprises must establish strict access and regulatory systems.

Lastly, conducting interdisciplinary research and promoting AI for good are essential. LLMs technology involves various fields such as computer science, natural language processing, cognitive science, ethics, and sustainable science. Comprehensive and in-depth research on LLMs requires multidisciplinary cooperation and collective wisdom. Researchers should comprehensively utilize theoretical perspectives and research methods from different disciplines, endeavoring to overcome disciplinary barriers, and explore the application prospects, risks, and response strategies of LLMs in assisting SDGs. For example, sustainable science helps define the sustainable development principles LLMs should follow (Deshmukh et al., 2010); cognitive science can reveal the cognitive mechanisms of LLMs, laying a foundation for human-machine collaboration; and ethics regulate the values choices of LLMs from a moral philosophical perspective. Only through interdisciplinary integration can LLMs be comprehensively understood, and AI systems with strong applicability and controllable risks be developed. Against the backdrop of rapid development in AI, it is imperative to strengthen research on AI ethics and governance, promote AI for good, and better serve SDGs. Policy-making departments should strengthen top-level design, promptly formulate governance principles and behavioral norms for LLMs and even the entire field of AI, strengthen accountability and regulatory mechanisms, and provide solid institutional guarantees for the responsible development and utilization of LLMs (Sadiq et al., 2023). Meanwhile, governments should increase support for basic research in AI, improve scientific fund allocation mechanisms, create an open, inclusive, and sustainable innovation ecosystem for AI, guide AI to contribute to positive areas such as education, medical progress, poverty reduction, and alleviate human development challenges with technology, continuously injecting new impetus into SDGs.

LLMs bring unprecedented opportunities to assist in achieving SDGs, but how to harness their capabilities and mitigate their risks requires continuous improvement in model training, value guidance, human-computer collaboration, and support from interdisciplinary research and responsible AI governance. It is essential to establish mechanisms for collaboration among governments, industries, academia, and research institutions to develop and utilize responsible LLMs, thereby helping to achieve the 2030 Agenda for Sustainable Development as scheduled.